\documentclass[conference]{IEEEtran}
\IEEEoverridecommandlockouts
\usepackage{cite}
\usepackage{amsmath,amssymb,amsfonts}
\usepackage{graphicx}
\usepackage{textcomp}
\usepackage{xcolor}

\usepackage{graphicx, caption, subcaption}
\usepackage{algpseudocode}
\usepackage{algorithm}
\usepackage{url}
 \usepackage{dblfloatfix}

\def\BibTeX{{\rm B\kern-.05em{\sc i\kern-.025em b}\kern-.08em
    T\kern-.1667em\lower.7ex\hbox{E}\kern-.125emX}}
\begin{document}

\title{Route-Forcing: Scalable Quantum Circuit Mapping for Scalable Quantum Computing Architectures
}

\makeatletter
\newcommand{\linebreakand}{%
  \end{@IEEEauthorhalign}
  \hfill\mbox{}\par
  \mbox{}\hfill\begin{@IEEEauthorhalign}
}
\makeatother

\author{\IEEEauthorblockN{Pau Escofet}
\IEEEauthorblockA{
\textit{Universitat Politècnica de Catalunya}\\
Barcelona, Spain \\
pau.escofet@upc.edu}
\and
\IEEEauthorblockN{Alejandro Gonzalvo}
\IEEEauthorblockA{
\textit{Universitat Politècnica de València}\\
València, Spain \\
agonhid@etsinf.upv.es}
\and
\IEEEauthorblockN{Eduard Alarcón}
\IEEEauthorblockA{
\textit{Universitat Politècnica de Catalunya}\\
Barcelona, Spain \\
eduard.alarcon@upc.edu}
\linebreakand
\IEEEauthorblockN{Carmen G. Almudéver}
\IEEEauthorblockA{
\textit{Universitat Politècnica de València}\\
València, Spain \\
cargara2@disca.upv.es}
\and
\IEEEauthorblockN{Sergi Abadal}
\IEEEauthorblockA{
\textit{Universitat Politècnica de Catalunya}\\
Barcelona, Spain \\
abadal@ac.upc.edu}
}

\maketitle

\begin{abstract}
Quantum computers are expected to scale in size to close the gap that currently exists between quantum algorithms and quantum hardware. To this end, quantum compilation techniques must scale along with the hardware constraints, shifting the current paradigm of obtaining an optimal compilation to relying on heuristics that allow for a fast solution, even though the quality of such a solution may not be optimal. Significant concerns arise as the execution time of current mapping techniques experiences a notable increase when applied to quantum computers with a high number of qubits. In this work, we present Route-Forcing, a quantum circuit mapping algorithm that shows a compilation time average speedup of $3.7\times$ compared to the state-of-the-art scalable techniques, reducing the depth of the mapped circuit by $4.7 \times$ at the expense of adding $1.3 \times$ more \texttt{SWAP} gates. Moreover, the proposed mapping algorithm is adapted and tuned for what is expected to be the next generation of quantum computers, in which different processors are interconnected to increase the total number of qubits, allowing for more complex computations.
\end{abstract}


\section{Introduction}
\label{sec:intro}
Quantum computing has emerged as a novel computational paradigm, leveraging the distinctive properties of quantum mechanics, such as superposition and entanglement \cite{nielsen_chuang_2010}. It allows the execution of specific calculations at an unprecedented speed, addressing challenges that are intractable for classical computers. The potential applications of quantum computing span diverse domains, ranging from cryptography, exemplified by Shor's prime factorization algorithm \cite{shor_polynomial_1997}, to optimized database searches, facilitated by Grover's algorithm \cite{grover_fast_1996}, or the simulation of physical systems \cite{Low2019hamiltonian}.

Despite the vast promise of quantum computing, a substantial gap exists between its potential and the current practical realization. Nowadays, quantum computers employ various qubit technologies, including superconducting qubits \cite{Nakamura_1999, RevModPhys.93.025005}, photonic qubits \cite{kok_linear_2007, srivastava_2015_optically}, quantum dots \cite{imagog_quantum_1999, 10.1063/1.5115814}, and trapped ions \cite{cirac_quantum_1995, Blatt2012} among others. However, regardless of the qubit technology used, contemporary quantum computers are limited up to a thousand qubits \cite{chow_2021_ibm, gambetta_2023_the} due to the challenges related to the integration of control circuits \cite{NAP25196} and the higher rate of undesirable qubit interactions leading to issues like crosstalk \cite{ding_systematic_2020}, falling considerably short of the million-qubit scale necessary for implementing fault-tolerant protocols and addressing real-world problems \cite{preskill_quantum_2018}.

For quantum computing architectures to scale, we need not only technologies that can allow the scaling of the number of qubits but also algorithms and compilation techniques that scale with such an increase in the number of qubits. One of the main constraints of current quantum processors, which is also expected to be present in future larger and scalable quantum chips, is the limited qubit connectivity. This restricts the possible interactions between qubits to, for instance, nearest-neighbour coupling. Consequently, quantum algorithms must undergo a compilation process known as quantum circuit mapping prior to their execution on a given quantum device. In this paper, we focus on quantum circuit mapping and how it needs to be improved for future quantum computing architectures.

\begin{figure*}[b]
\vspace{-0.5cm}
\centering
\begin{subfigure}{0.13\textwidth}
    \raisebox{0.45cm}{\includegraphics[width=\textwidth]{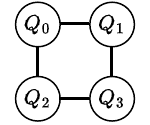}}
    \vspace{-0.8cm}
    \caption{Processor's coupling map}
    \label{fig:qc_topology}
\end{subfigure}
\hfill
\begin{subfigure}{0.225\textwidth}
     \raisebox{0.15cm}{\includegraphics[width=\textwidth]{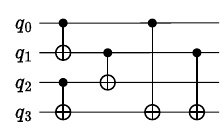}}
    \vspace{-0.8cm}
    \caption{Quantum circuit with 4 qubits and 5 two-qubit gates}
    \label{fig:quantum_circuit}
\end{subfigure}
\hfill
\begin{subfigure}{0.275\textwidth}
    \includegraphics[width=\textwidth]{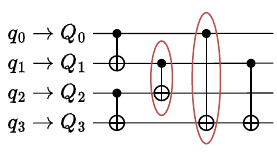}
    \vspace{-0.8cm}
    \caption{Virtual qubits mapped to physical qubits. Unfeasible gates are circled.}
    \label{fig:initial_placement_example}
\end{subfigure}
\hfill
\begin{subfigure}{0.33\textwidth}
    \raisebox{0.15cm}{\includegraphics[width=\textwidth]{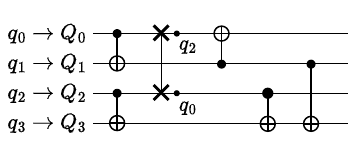}}
    \vspace{-0.8cm}
    \caption{Final mapped circuit. A \texttt{SWAP} gate ($\times$) is added to satisfy the connectivity constraint}
    \label{fig:mapped_circuit}
\end{subfigure}
\caption{Overview of the process of mapping a quantum circuit into the topology of a particular quantum computer.}
\label{fig:monolithic_mapping}
\end{figure*}

This paper introduces Route-Forcing, a novel quantum circuit mapping algorithm designed to address not only the challenges posed by limited qubit connectivity but also to obtain a low compilation time, enabling the algorithm to be used for larger architectures. The algorithm considers attraction forces among qubits that need to interact bringing them closer together. Multiple interactions can be considered simultaneously, adding parallel \texttt{SWAP} gates that help maintain a low depth for the generated circuit. Moreover, the proposed algorithm is easily adaptable to the needs of future, modular, scalable quantum computers, a new architectural paradigm that connects several small quantum processors to obtain the computational power of a larger one \cite{9268630}.

The major contributions of this paper can be summarized as follows:
\begin{itemize}
    \item We provide the insights needed to design quantum circuit mapping techniques for the incoming quantum computing architectures.
    \item We propose a novel quantum routing algorithm that focuses on scalable architectures, taking into account the fidelity of each added operation, aiming to maximize the success probability of the mapped algorithm.
    \item We perform a design space exploration to understand the relation between the algorithm's hyperparameters and the obtained solution, providing insights on how to modify the algorithm's parameters to prioritize one metric over the others.
    \item Lastly, we compare the proposed approach with the state-of-the-art scalable quantum mapping techniques, obtaining an average speedup in the compilation time of $3.7 \times$, reducing the depth of the mapped circuit by $4.7 \times$ at the expense of adding $1.3 \times$ more \texttt{SWAP} gates.
\end{itemize}

The rest of the paper is organized as follows. In Section \ref{sec:background}, we provide the essential background on quantum circuit mapping and quantum computing architectures needed to understand this work, with a particular focus on the scalability of quantum processors. Our solution is proposed and described in Section \ref{sec:mapping_by_forces}, accompanied by a design space exploration to understand how different hyperparameters impact quantum circuit mapping. Section \ref{sec:experiments_results} delves into the algorithm's performance in single-core and multi-core quantum computers. Lastly, Section \ref{sec:related_work} summarizes related work, and Section \ref{sec:conclusions} concludes the paper by outlining future directions for this research.

\section{Background}
\label{sec:background}

Quantum algorithms are usually represented as quantum circuits (Figure \ref{fig:quantum_circuit}), where quantum gates are used to modify the quantum states. For a quantum circuit to be successfully executed on a quantum computer, it must align with the inherent constraints and limitations imposed by the architecture's topology (Figure \ref{fig:qc_topology}). This entails a meticulous adjustment of interactions within the circuit to conform to the specific connectivity patterns and restrictions dictated by the quantum processor's hardware. In essence, the quantum circuit must undergo a tailored transformation phase, ensuring that every two-qubit gate from the circuit is executed on two neighbouring physical qubits (i.e. qubits that can interact). Such transformation stage is depicted in Figure \ref{fig:monolithic_mapping} throughout a simple example.

This problem is known as \emph{quantum circuit mapping} and comprises two different processes, summarized as follows. Firstly, each virtual (or logical) qubit from the circuit is allocated into a physical qubit of the processor throughout a one-to-one mapping (Figure \ref{fig:initial_placement_example}). This stage is called initial placement or qubit allocation, and its aim is to place interacting qubits in neighbouring physical qubits such that most, if not all, two-qubit gates can be executed. However, there still might be some two-qubit gates from the circuit that may involve not neighbouring qubits (see circled CNOT gates in Figure \ref{fig:initial_placement_example}). In this case, \texttt{SWAP} gates will be added to the initial circuit to move quantum states into neighbouring physical qubits, as depicted in Figure \ref{fig:mapped_circuit}. This second process is known as the routing of qubits. Mapping a quantum circuit to a restricted-connectivity quantum processor usually results in a gate overhead that increases the execution time of the algorithm, which, in turn, decreases its success rate.

For a comprehensive understanding of the current state-of-the-art techniques in quantum circuit mapping, readers are directed to Section \ref{sec:related_work}, where an explanation of related work is provided.


\subsection{Mapping in Monolithic Architectures}
\label{sec:single-core_mapping}

The mapping problem for single-core quantum processors has been well-studied, and several quantum circuit mapping algorithms have been proposed \cite{lao_2022_timing, 8342181, 9643554, 9923822, li_2019_tackling, 10.1145/3445814.3446706}, both optimal and based on heuristics. However, this problem has been proved to be NP-Complete \cite{qubit_allocation}, which makes those mapping algorithms that find the optimal solution not suitable for compiling large-scale algorithms due to an excessive compilation time (time for the mapping algorithm to find the solution).

So far, most of the proposed quantum circuit mapping techniques have been evaluated based on the number of added \texttt{SWAPs} (i.e. gate overhead) and on the resulting circuit depth/latency, aiming at minimizing both metrics. Note that the circuit depth is defined as the length of the critical path in the circuit, whereas circuit latency considers the gate duration. However, compilation time is becoming a critical parameter as the number of qubits increases. The SABRE algorithm, proposed in \cite{li_2019_tackling}, is the first quantum circuit mapping technique that considers scalability and will serve as a baseline for this work.


\subsection{Scalable Quantum Computing Architectures}


One of the main challenges of quantum processors is scalability (i.e. increasing the number of qubits). It is hard to scale up current monolithic single-chip quantum architectures, primarily due to the intricate integration challenges associated with control circuitry and wiring for qubit access, all while striving to uphold low error rates \cite{NAP25196}. In addition, increasing the number of qubits within a single processor comes with issues like undesirable qubit interactions, such as crosstalk \cite{ding_systematic_2020}. Consequently, the pursuit of scaling monolithic quantum computers to accommodate higher qubit counts stands as a formidable challenge, necessitating innovative strategies to overcome this bottleneck.

A promising solution is found when interconnecting several quantum processors, leading to modular or multi-core quantum computing architectures \cite{Bravyi_2022, jnane_multicore_2022, rodrigo_exploring_2020, smith_scaling_2022, laracuente_modeling_2023, alarcon_scalable_2023, ang2022architectures}. Quantum cores are connected through quantum-coherent and/or classical links \cite{gold_entanglement_2021, smith_scaling_2022}, allowing for the transfer of quantum states across cores. Adopting a modular architecture presents a viable solution to the scalability challenge while preserving quantum coherence.

Transitioning from monolithic to multi-core quantum processors requires the development of a new breed of compilation method, whose main aim is minimizing the number of inter-core communications. Communication between cores in multi-core quantum processors incurs significantly higher costs than intra-core communications (i.e. \texttt{SWAPs}), posing optimization challenges of considerable complexity. Therefore, the quantum circuit mapping algorithm must be tailored to match this scalable architectural paradigm.

\subsection{Mapping in Multi-Core Architectures}
\label{sesc:multi-core_mapping}

Different levels of modularity, each of them with specific interconnections, are envisioned for scaling up quantum computers in the next years \cite{Bravyi_2022}.

The mapping problem for multi-core architectures highly depends on how cores are interconnected \cite{Escofet2023InterconnectFF}. If the inter-core links allow for the coupling of qubits \cite{gold_entanglement_2021, smith_scaling_2022}, the connectivity among qubits resembles that of a single-core processor, with the difference that the inter-core connections will have higher error rates. On the other hand, for those architectures that base the inter-core data transfer in EPR communication primitives \cite{rodrigo_exploring_2020}, a new abstraction must be made since the generation of entangled pairs is costly and will act as a bottleneck.

Nevertheless, in both modular architecture paradigms, the mapper needs to be designed with two main goals:
\begin{enumerate}
    \item The number of non-local communications needs to be minimized, as the inter-core links will have less fidelity and/or a higher latency.
    \item Since modular architectures will host a higher number of qubits, the mapping needs to be efficient, obtaining a solution in a relatively small amount of time.
\end{enumerate}

Several works have been recently published regarding the mapping of quantum circuits into modular architectures using EPR-based communications \cite{baker_time-sliced_2020, bandic_mapping_2023, escofet_hungarian_2023, escofet_revisiting_2024, wu_collcomm_2022, sundaram_distribution_2022}, focusing on the minimization of non-local communications, and the compilation time. In this work, we focus on the mapping problem for multi-core architectures but with short-range coupling links between cores, as proposed and proved in \cite{gold_entanglement_2021}, and depicted in Figure \ref{fig:multi-core}. For this type of multi-core architecture, a single-core mapper would also work as the qubit connectivity resembles the one from a single-core quantum processor. However, most single-core quantum mappers do not scale well when increasing the number of qubits, making them useless in this scenario, except for \cite{li_2019_tackling}, where Li \textit{et al.} proposed the first scalable mapping algorithm.

\begin{figure}[t]
\centering
\includegraphics[width=0.8\columnwidth]{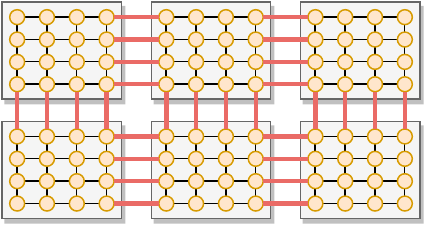}
\caption{Multi-core architecture with short-range coupling links between cores. Black edges represent intra-core links with high fidelity, and red edges represent inter-core links with lower fidelity.}
\label{fig:multi-core}
\end{figure}

\section{Route-Forcing: Attracting Qubits}
\label{sec:mapping_by_forces}

In this section, we explain \emph{Route-Forcing}, our proposed algorithm for routing quantum circuits into a scalable restricted topology. The design goals for this algorithm are the following, as previously stated in Section \ref{sesc:multi-core_mapping}: 1) being able to avoid low-fidelity edges, and 2) obtaining a valid solution within a low execution time without compromising accuracy. Though mapping is usually composed of two stages, initial placement and routing, in this work, we only target the latter one.

The routing algorithm's inputs are the processor's topology (qubits as nodes and edges representing the physical coupling link between qubits), the quantum circuit (composed of quantum gates applied to virtual qubits), and a set of hyperparameters that will be introduced later.

After decomposing the circuit into single and two-qubit gates, it is transformed into an ordered set of operations, applying a topological sort on the quantum gates. By doing so, we obtain a Directed Acyclic Graph (DAG) $Op$, representing the execution constraints between the different gates in the circuit. Such data structure is illustrated in Figure \ref{fig:operations_topological_sort}.

\begin{figure}[t]
\centering
\includegraphics[width=\columnwidth]{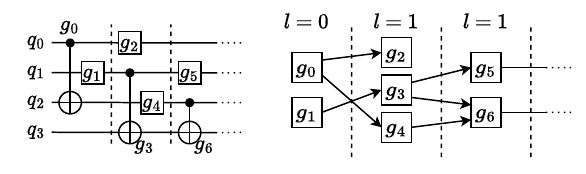}
\caption{Directed acyclic graph of the operations of the circuit. For an operation to be executed, all predecessors need to have been executed already.}
\label{fig:operations_topological_sort}
\end{figure}

In the circuit's DAG, those operations with no incoming edges (gates in the first layer of Figure \ref{fig:operations_topological_sort}) are the executable ones, as there are no predecessors to be executed. Once a gate is performed, it is removed from the DAG, unlocking other gates to be executed. This data structure will be modified along the circuit mapping process, ensuring no gate is executed before its predecessors.

Table \ref{tab:notation} shows the notation used in subsequent sections.

\begin{table}
    \centering
    \caption{Notation used in this work.}
    \begin{tabular}{cc}
        \textbf{Notation} & \textbf{Definition} \\
         $q_{1, ..., n}$& Virtual Qubits from the Quantum Circuit\\
         $Q_{1, ..., m}$& Physical Qubits from the Architecture\\
         $G(Q,E)$ & Coupling graph of the chip\\
         $Op_{l}$ & Layer $l$ of the Operations DAG $Op$\\
         $\pi$ & Allocation of virtual to physical qubits\\
         $\pi_{q_A}$ & Physical allocation of virtual qubit $q_A$ in $\pi$\\
    \end{tabular}
    \label{tab:notation}
\end{table}

\subsection{Routing of Qubits}
\label{sec:routing}

After obtaining the initial placement $\pi'$, the task of the routing algorithm is to ensure that, whenever two qubits need to interact, they are placed in neighbouring physical qubits.

The result will be a new quantum circuit with the same functionality as the original one but with some $\texttt{SWAP}$ operations added, which are required to move virtual qubits from one physical qubit to another along the execution of the circuit to cope with the connectivity constraints of the processor.

The proposed algorithm is summarized in Algorithm \ref{alg:routing}.

\begin{algorithm}
\caption{Route-Forcing Algorithm}
\label{alg:routing}
\begin{algorithmic}[0]
    \Require Processor's topology $G=(Q,E)$, Operations DAG $Op$, Initial placement $\pi'$, Lookahead value $k$
    \Ensure Augmented circuit $QC$
    \State $QC \gets \texttt{Empty Circuit}$
    \State $\pi \gets \pi'$
    \While{$Op$ not Empty}
        \State $F = Op_0$ 
        \For{gate $g \in F$}
            \If{$g$ executable in $\pi$}
                \State $QC.\texttt{add}(g)$
                \State $Op.\texttt{remove}(g)$
            \EndIf
        \EndFor
    
        \For{Layer $l \leq k$}
            \For{Gate $g \in Op_l$}
                \State $q_A, q_B \gets g.\texttt{qubits()}$
                \State $f_{q_A} \gets \pi_{q_B} - \pi_{q_A}$
                \State $f_{q_B} \gets \pi_{q_A} - \pi_{q_B}$
        
                \For{$e \in G(\pi_{q_A})$}
                    \State $\texttt{SWAP}_e += f_{q_A} \cdot e \cdot d(l)$
                \EndFor
        
                \For{$e \in G(\pi_{q_B})$}
                    \State $\texttt{SWAP}_e += f_{q_B} \cdot e \cdot d(l)$
                \EndFor
            \EndFor
        \EndFor
        \State $\texttt{SWAP}.\texttt{sort()}$
        \For{Executable $\texttt{SWAP}_e \in \texttt{SWAP}$}
            \State $\pi.\texttt{update}(\texttt{SWAP}_e)$
        \EndFor
    \EndWhile
\end{algorithmic}
\end{algorithm}

To add the necessary gates for qubit routing, we will start with the initial placement $\pi'$, previously computed. Then, traversing the circuit's DAG, an attraction force will be set for those qubits that need to interact, and, based on this attraction force and the immediacy of the interaction each force represents, a $\texttt{SWAP}$ coefficient will be computed for each possible edge.

At any time step on the mapping process $\pi$, we know the location of a virtual qubit $q_A$, meaning the $x$ and $y$ coordinates of the physical qubit currently holding its value ($\pi_{q_A}$). Therefore, if the virtual qubit $q_A$ needs to interact with the virtual qubit $q_B$, we set attraction forces $\overrightarrow{f_{q_A}}$ and $\overrightarrow{f_{q_B}}$ between both qubits as:

\begin{equation}
    \overrightarrow{f_{q_A}} = \overrightarrow{(\pi_{q_B} - \pi_{q_A})} \qquad \qquad \overrightarrow{f_{q_B}} = \overrightarrow{(\pi_{q_A} - \pi_{q_B})}
\label{eq:attraction_1}
\end{equation}

Note that both $\overrightarrow{f_{q_A}}$ and $\overrightarrow{f_{q_B}}$ are two-position vectors, with $x$ and $y$ coordinates. Note also that $\overrightarrow{f_{q_A}} = - \overrightarrow{f_{q_B}}$, since the attraction force is computed as the \textit{final} position minus the \textit{initial} (or \textit{current}) position.

It is important to differentiate interactions from the first layer of the circuit's DAG (all their predecessors have been executed) from those that still have predecessors to execute. For this, we assign a value $l$ to each interaction of the DAG based on how many layers of operations need to be executed to reach that operation. Therefore, the operations at the head will have $l=0$, those unlocked by executing only the operations in the head will have $l=1$, and so on.

These layer levels will help us compute the $\texttt{SWAP}$ coefficient for each edge by using an exponential decay function $w(l)$ to give more strength to early interactions,

\begin{equation}
    w(l) = 2^{-l}.
\label{eq:decaying_1}
\end{equation}

Therefore, when considering how to move a qubit $q$, we will first compute the $\texttt{SWAP}$ coefficient of all the edges connecting $\pi_{q}$ as the dot product between the attraction force vector, and the vector describing the edge $\overrightarrow{e_{q,i}} = \overrightarrow{(Q_i - \pi_q)}$, where physical qubit $Q_i$ is connected with the physical qubit holding the virtual qubit $q$ (i.e. $\pi_q$).

Then, the $\texttt{SWAP}$ coefficient of an edge is a scalar value computed as:
\begin{equation}
    \texttt{SWAP}_i = \overrightarrow{f_q} \cdot \overrightarrow{(Q_i - \pi_q)} \cdot w(l) = \overrightarrow{f_q} \cdot \overrightarrow{e_{q,i}} \cdot w(l)
\label{eq:first_swap}
\end{equation}

The greater the $\texttt{SWAP}$ coefficient is, the greater the improvement of performing such $\texttt{SWAP}$. The coefficient can also be negative, meaning that performing the $\texttt{SWAP}$ along that edge will worsen the placement for the upcoming quantum gates.

Figure \ref{fig:attraction-example} depicts an example of the $\texttt{SWAP}$ coefficient for the four edges connecting the blue qubit, computed using Equation (\ref{eq:first_swap}), and considering just one layer of the DAG.

\begin{figure}[t]
\centering
\includegraphics[width=0.9\columnwidth]{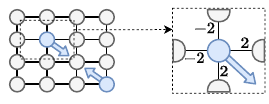}
\caption{Attraction forces and \texttt{SWAP} coefficient example.}
\label{fig:attraction-example}
\end{figure}

When considering multiple layers of the DAG, $\texttt{SWAP}$ coefficients of the same edge will be added, strengthening a possible $\texttt{SWAP}$ when having multiple interactions in the same direction and weakening it otherwise. We encode the number of layers from the DAG to be considered as a hyperparameter of the algorithm, the $k$ value. When setting a value of $k=0$, only the interactions in the first layer of the DAG will be considered.

Once all qubit interactions have been considered, we sort the edges from the processor's topology by descending order of their $\texttt{SWAP}$ coefficients and then, we greedily apply $\texttt{SWAP}$ operations. When performing a $\texttt{SWAP}$ operation, the other edges connecting the involved physical qubits will no longer be candidates, as the virtual qubits held in the physical ones have changed.

\subsection{Modelling the Attraction Forces}
\label{sec:modelling_attraction_forces}

We have now proposed one equation for computing the $\texttt{SWAP}$ coefficients for the edges, using Equation (\ref{eq:attraction_1}) for the attraction force and Equation (\ref{eq:decaying_1}) for the interactions weighting factor. As it is now, the value of the attraction force depends on the distance of the two interacting qubits, being upper-bounded by the size of the architecture. However, the interaction weighting factor does not depend on the architecture size.

The question of whether the equations should be agnostic of the size of the architecture remains open. Nevertheless, if the attraction force is formulated based on the topology's dimensions, the interaction weighting factor must also be modelled accordingly. Figure \ref{fig:attraction-weighting-codesign} exemplifies a (1-dimension) case where the green qubits interact at level 0 of the operation's DAG and, since they are close together, their respective interactions in deeper layers of the DAG drive them further apart from each other. This exemplifies why the attraction force $\overrightarrow{f_{q}}$ and the interaction weighting factor $w(l)$ should be defined together.

\begin{figure}[t]
\centering
\includegraphics[width=\columnwidth]{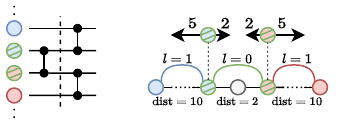}
\caption{The attraction force function and the interaction weighting factor must be defined together to match the architecture. In this simple example, the two green qubits will be moved far apart, even though their interaction is at the head of the operations DAG. This counter-intuitive scenario is due to the distance of the second interaction of each green qubit (blue and red, respectively).}
\label{fig:attraction-weighting-codesign}
\end{figure}

To this end, we propose to modify the interaction weighting function and make it dependent on the architectural size. Since we want no interaction in layer $l+1$ overruling an interaction in layer $l$. We use the exponential decaying function, with the diameter $d$ of the topology as the decaying ratio. The diameter $d$ of a graph is defined as the length of the shortest path between the most distanced nodes (i.e. the longest shortest path in the graph). Therefore, we will use the following function to weigh the interactions at layer $l$ of the circuit's DAG.
\begin{equation}
    w_d(l) = d^{-l}
\end{equation}

With this, the proposed policy to compute the $\texttt{SWAP}$ coefficients results as follows:
\begin{equation}
    \texttt{SWAP}_i = \overrightarrow{f_q} \cdot \overrightarrow{e_{q,i}} \cdot w_d(l) = \overrightarrow{(\pi_{q'} - \pi_{q})} \cdot \overrightarrow{e_{q,i}} \cdot d^{-l}
\label{eq:swap_policy_1}
\end{equation}

Moreover, aiming at not adding unnecessary $\texttt{SWAPs}$, another hyperparameter of the algorithm will be a threshold on the $\texttt{SWAP}$ coefficient called the \textit{penalization SWAP parameter} $p$. The edges with $\texttt{SWAP}$ coefficient below this threshold (i.e. $\texttt{SWAP}_i < p$) will not be considered. 

\subsection{Considering the link's fidelities}
\label{sec:considering_link_fidelity}

Multi-core quantum computing architectures \cite{Bravyi_2022, jnane_multicore_2022, rodrigo_exploring_2020, smith_scaling_2022, laracuente_modeling_2023, alarcon_scalable_2023} are envisioned to host a higher number of qubits than monolithic architectures, enabling the execution of more extensive quantum circuits or incorporating quantum error correction codes \cite{7336474}. In this section, we extend the Route-Forcing algorithm to consider low-fidelity, inter-core links \cite{gold_entanglement_2021}. Moreover, as this extension considers the fidelity of each quantum link in the architecture, it can be used whenever the architecture presents heterogeneity on the link fidelities \cite{tannu_not_2019}.

We now want to prioritize those $\texttt{SWAP}$ gates occurring across high-fidelity edges rather than those occurring through inter-core edges. To do so, we will use the edge fidelity $F$ of each link in the architecture. The $\texttt{SWAP}$ coefficients will be scaled by a power of the edge fidelity. Therefore, a new hyperparameter $r$ is introduced, which will be used as the exponent for the fidelity, leading to a new $\texttt{SWAP}$ policy:

\begin{equation}
    \texttt{SWAP}_i = \overrightarrow{f_q} \cdot \overrightarrow{e_{q,i}} \cdot w_d(l) \cdot F_i^r
\label{eq:swap_policy_fidelity_1}
\end{equation}

The higher the $r$ value, the higher the impact of the edge's fidelity on the \texttt{SWAP} coefficient. If $r$ is set to zero, the algorithm does not take into account the edge's fidelity.

\subsection{On the Algorithm's Convergence}
\label{sec:convergence}

We have identified some particular scenarios where Route-Forcing may not converge, performing the same $\texttt{SWAPs}$ over and over. An example of such scenarios is depicted in Figure \ref{fig:critical_configurations}, where two interactions exist in the same layer, one between green qubits and the other between blue qubits. The $\texttt{SWAP}$ coefficients shown in the edges have been computed using Equation (\ref{eq:swap_policy_1}). The red edges are the ones the algorithm selects, thus exchanging the positions of the blue and green qubits, which will lead to the same configuration (and thus the same $\texttt{SWAP}$ coefficients) but with the green and blue qubits swapped. In this example, all link's fidelities are set to 1.

\begin{figure}[t]
\centering
\includegraphics[width=0.5\columnwidth]{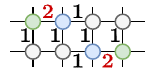}
\caption{Critical configuration for the $\texttt{SWAP}$ policy, disrupting the algorithm's convergence.}
\label{fig:critical_configurations}
\end{figure}

To solve this problem with the minimum possible time overhead, we will introduce some randomness when selecting on which edges we should apply a $\texttt{SWAP}$ operation. After sorting the edges by its $\texttt{SWAP}$ coefficient, we will randomly exchange the position of some edges in the ordered list. By doing this, the algorithm will likely increase the number of $\texttt{SWAPs}$ added, but the non-convergence of the algorithm will be avoided.

\subsection{Tuning the Algorithm: a Design Space Exploration}
\label{sec:dse}

\begin{figure*}
\centering
\begin{subfigure}[t]{0.475\textwidth}
    \includegraphics[width=\textwidth]{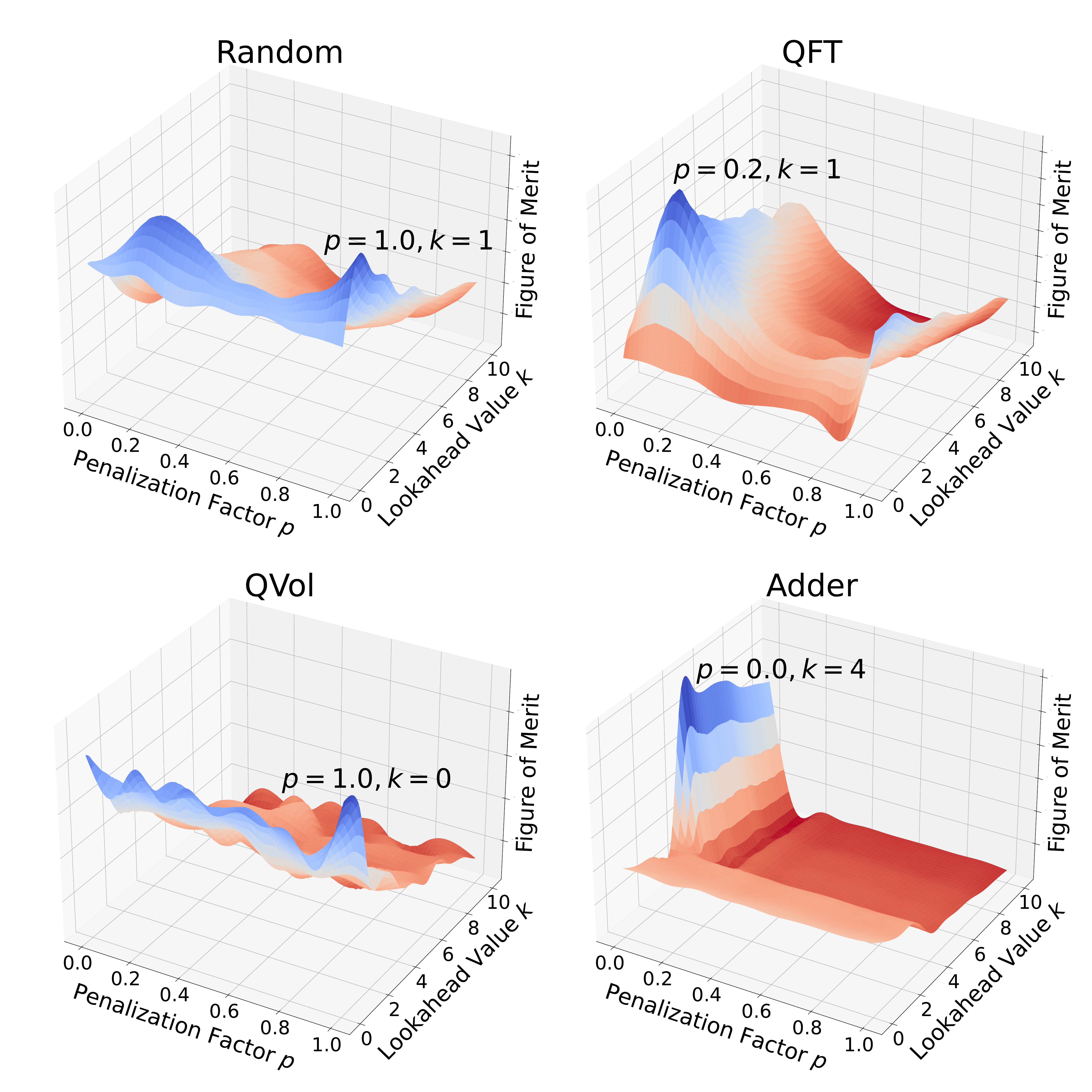}
\end{subfigure}
\hfill
\begin{subfigure}[t]{0.475\textwidth}
    \includegraphics[width=\textwidth]{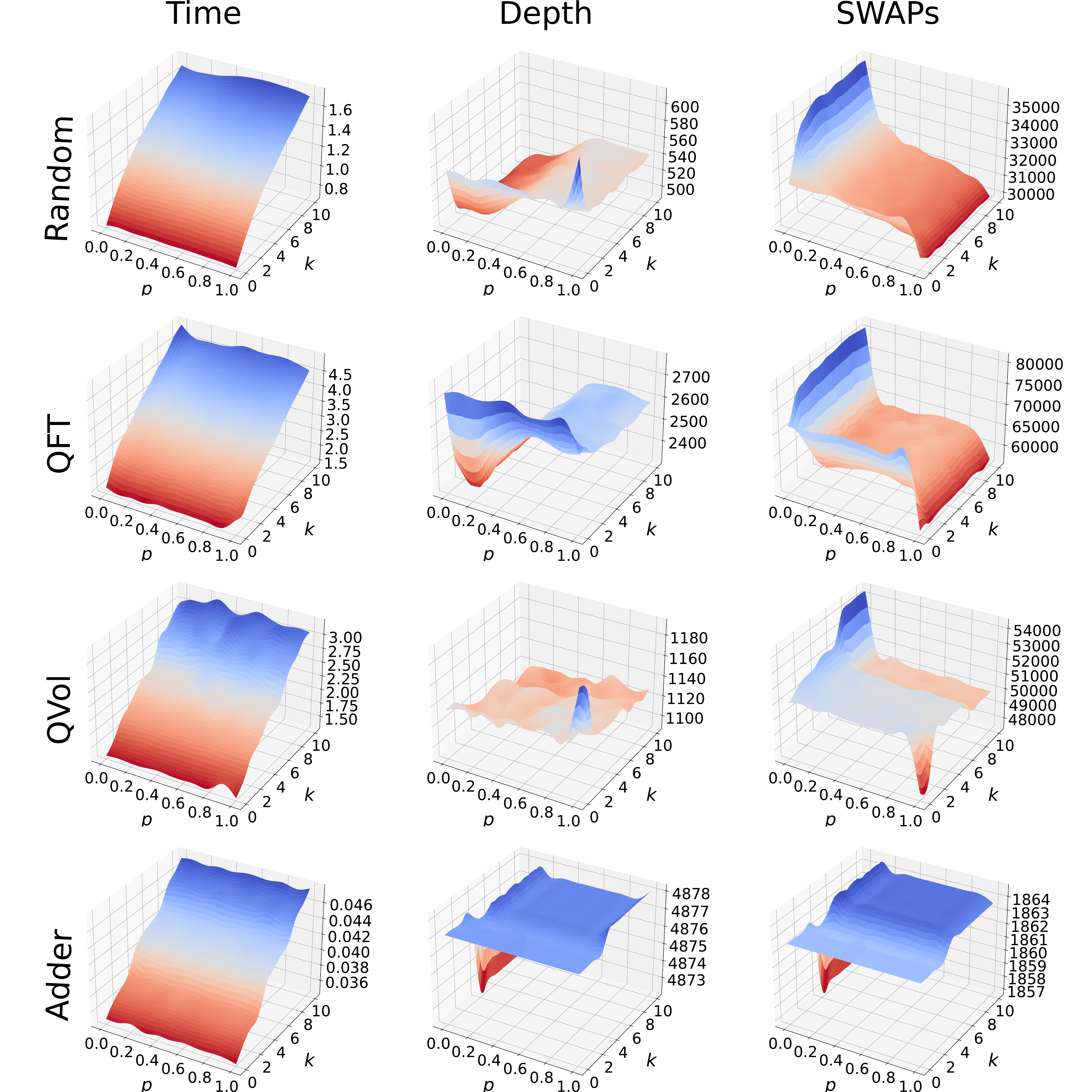}
\end{subfigure}
\caption{Design space exploration for a quantum computing architecture with 256 qubits in a grid topology. Each evaluated metric (execution time, final circuit depth, and number of $SWAP$ gates added) is used for the benchmark-wise figure of merit (left). The optimal configurations for each quantum algorithm are as follows: Random Circuits $p=1.0, k=1$, Quantum Fourier Transform $p=0.2, k=1$, Quantum Volume $p=1.0, k=0$, and Cuccaro Adder $p=0.0, k=4$. Results are averaged over 50 trials.}
\label{fig:DSE}
\end{figure*}

Before comparing our algorithm to state-of-the-art quantum circuit mappers, we explore the effects of the lookahead hyperparameter $k$ and the penalization factor $p$ to understand the relationship between the mapping algorithm and the circuit's structure. To do so, we measure the number of $\texttt{SWAPs}$ added by the mapping algorithm, the circuit depth after the mapping, and the execution time of the mapping algorithm, aiming at minimizing each considered metric.

We will test the algorithm using four different quantum circuits: Random Quantum Circuit, Quantum Fourier Transform \cite{nielsen_chuang_2010}, Quantum Volume \cite{cross_2019_validating}, and Cuccaro Adder \cite{cuccaro2004new}. We have selected these benchmarks as they comprise different types of quantum circuits. The QFT and Adder circuits are structured and organized, while the Random Circuits lack structure. Lastly, Quantum Volume is a well-established benchmarking algorithm for quantum computers. All circuits have been obtained from Qiskit's library \cite{Qiskit}.

We will map each quantum circuit to a $16 \times 16$ grid topology, totalling $256$ qubits, a number large enough to see trends in the metrics but small enough to allow for an in-depth hyperparameter search. As it leads to a low execution time, many hyperparameter combinations can be tested.

We will use integer values from $0$ to $10$ for the lookahead $k$ and values from $0$ to $1.0$, with a step of $0.1$ for the penalization factor $p$. The three evaluation metrics used (execution time, depth, and \texttt{SWAPs} added) are combined into a figure of merit after being normalized, forcing each independent metric to have a value between 1 and 2. The figure of merit is:

\begin{equation}
    \text{Fig. of Merit} = \frac{1}{\text{Time} \cdot \text{Depth} \cdot \text{SWAPs}}
\end{equation}

All experimental procedures were conducted on a computing system featuring an AMD Ryzen 9 5950X @ 3.7GHz, equipped with 128 GB of RAM and 16 cores, operating on Ubuntu 20.04. The simulation procedure was implemented utilizing Python 3.8. All circuits have been obtained from Qiskit's library \cite{Qiskit}. Each combination (circuit, lookahead value $k$, and penalization factor $p$) has been executed 50 times with a different random initial placement for each iteration. Results have been averaged across repetitions.

Figure \ref{fig:DSE} shows the figure of merit, with the optimal hyperparameter combination for each quantum algorithm tested, as well as each independent metric for each combination of lookahead value and penalization factor. 

It can be seen how, irrespective of the quantum circuit, using a higher $k$ leads to higher execution time. This is an expected result since more computations are needed at each step of the iterative mapping algorithm. A higher $p$ value usually implies a lower number of $\texttt{SWAPs}$, but a higher circuit depth, which may seem counter-intuitive, as usually, the higher the number of gates, the higher the circuit depth is. It can also be seen how the penalization factor $p$ has no impact on the execution time but a huge impact on the number of \texttt{SWAP} gates added.

When $k$ increases, the circuit depth is reduced because of the parallelization of $\texttt{SWAPs}$. At each iteration of the mapping algorithm, all the $\texttt{SWAP}$ gates added can be executed in parallel, increasing the depth of the circuit by one. A higher $k$ value implies more interactions are considered at each iteration, adding more $\texttt{SWAPs}$. However, since the extra $\texttt{SWAPs}$ will be executed in parallel, they do not contribute to the depth of the final circuit.

One of the most insightful results in this exploration is that the Cuccaro Adder, a highly structured circuit, performs better for a higher $k$. In contrast, unstructured ones (Random and Quantum Volume) perform better for a lower $k$ since future interactions do not follow any specific pattern and are difficult to account for.

\section{Experiments and Results}
\label{sec:experiments_results}

This section evaluates the Route-Forcing mapping algorithm against a state-of-the-art scalable quantum circuit mapper. Though there are many algorithms for mapping quantum circuits into single-core algorithms \cite{lao_2022_timing, 8342181, 9643554, 9923822, li_2019_tackling}, we use the SABRE algorithm, introduced in \cite{li_2019_tackling}, as the state-of-the-art alternative, since it focuses on providing a scalable solution, rather than an optimal one, analogous to our work.

We have used Qiskit's \cite{Qiskit} implementation of SABRE. We employ solely the routing pass without further optimization through techniques such as gate optimization or other available passes from Qiskit. To assess only the routing method, both approaches will start with the same random initial placement, which will change at each trial. Moreover, the Route-Forcing algorithm has been coded in Python, matching the implementation of SABRE to ensure a fair comparison between both approaches.

We use two different SABRE strategies, \textit{basic} and \textit{decay}, where the \textit{decay} strategy is an extension of the \textit{basic} one where \texttt{SWAP} gates that can be executed in parallel are prioritized. We recommend SABRE's original paper \cite{li_2019_tackling} to understand the differences between both strategies.

\subsection{SWAP based comparison}
For the first experiment, we will map the same four quantum circuits as in the previous hyperparameter exploration into a grid topology with an increasing number of qubits, from 64 ($8 \times 8$ grid) to 1024 qubits ($32 \times 32$ grid), to understand how these mapping algorithms will perform when increasing the size of the quantum computing architecture.

Figure \ref{fig:mapping_monolithic} shows, for each quantum circuit, the ratio of execution time, depth, and \texttt{SWAPs} added when mapping using the SABRE algorithm or our proposal. The top row uses the \textit{basic} heuristic for SABRE, and the lower one uses the \textit{decay} heuristic. Each ratio is computed as SABRE performance divided by Route-Forcing performance, which means our approach outperforms SABRE when the ratio is above one.

\begin{figure*}
\centering
\includegraphics[width=\textwidth]{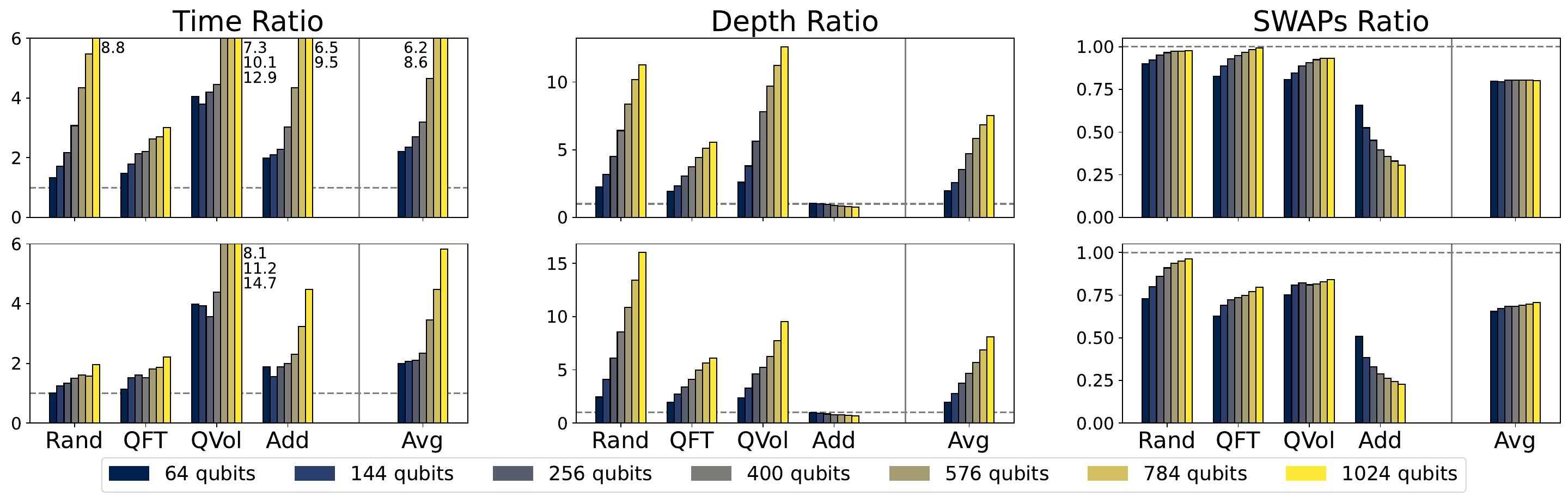}
\caption{Performance ratio (the higher, the better, since it is computed as SABRE/Route-Forcing) of the three evaluated metrics (execution time, circuit depth, and added gates) across all tested circuits (Random Circuits, Quantum Fourier Transform, Quantum Volume, and Cuccaro Adder), and all computer sizes (from 64 to 1024 qubits in a square grid topology). The top row uses the \textit{basic} heuristic, and the lower one uses the \textit{decay} heuristic. Results are averaged over 50 trials.}
\label{fig:mapping_monolithic}
\end{figure*}


We can see that, on average, for each circuit used, the proposed algorithm obtains a faster solution than the state-of-the-art considered, which was the primary goal of the work. Moreover, the circuit depth obtained with the Route-Forcing algorithm is much better for the Random Circuits, Quantum Volume and Quantum Fourier Transform but not for the Cuccaro Adder. Such improvement in the depth comes from adding multiple parallel \texttt{SWAP} gates, in \cite{li_2019_tackling}, just one \texttt{SWAP} gate is added at each iteration, thus reducing the total number of \texttt{SWAPs}, at the expense of a much higher circuit depth for most benchmarks. 

Based on the results obtained, it is noteworthy that the speedup in compilation time (time ratio) and depth ratio improves with increasing system size. This suggests that the advantages of employing Route-Forcing become more pronounced as quantum computing architectures scale up. Additionally, the \texttt{SWAP} ratio maintains a consistent average, indicating that transitioning to larger devices does not diminish Route-Forcing's performance relative to SABRE.


\subsection{Fidelity-based comparison}
We now repeat the assessment conducted earlier for single-core platforms, but mapping quantum circuits into architectures with variability in the link fidelity, using the re-defined $\texttt{SWAP}$ policy (Equation (\ref{eq:swap_policy_fidelity_1}).

For this comparison, we will assign a fidelity $F$ to each link in the architecture. By means of using the re-defined $\texttt{SWAP}$ policies, our algorithm should be able to avoid low-fidelity links and prioritize those links with higher fidelity.

We previously used the number of $\texttt{SWAPs}$ as an evaluation metric. However, in this scenario, not all $\texttt{SWAPs}$ have the same impact now that edges have different fidelities, and a solution containing less $\texttt{SWAPs}$ can perform worse than another one if many $\texttt{SWAPs}$ involve links with low fidelity.

Current methods for obtaining circuit fidelity rely on simulating the whole quantum state, which is out of the picture for systems as big as the ones we are using. Therefore, we will use the Estimated Success Probability (ESP), a metric that considers each gate's fidelity in the circuit \cite{10.1145/3386162, 10.1145/3297858.3304007}.

$$ESP = \prod_{g \in Op} F_g$$

Since SABRE does not consider the fidelity of the links to compute the routing, we begin by assessing the impact of considering the link fidelity when routing qubits for each selected benchmark only for the Route-Forcing algorithm. To do so, we compute the ESP for different $r$ values, starting from 0 (not considering link fidelity) and going up to 50.

Figure \ref{fig:fidelity_r} shows the normalized ESP (between 0 and 1) obtained for each considered circuit when mapping it for an increasing value of $r$. In the architecture, each link has a fidelity randomly selected from a uniform distribution ranging between $0.999$ and $0.9999$. 


\begin{figure}[t]
\centering
\includegraphics[width=\columnwidth]{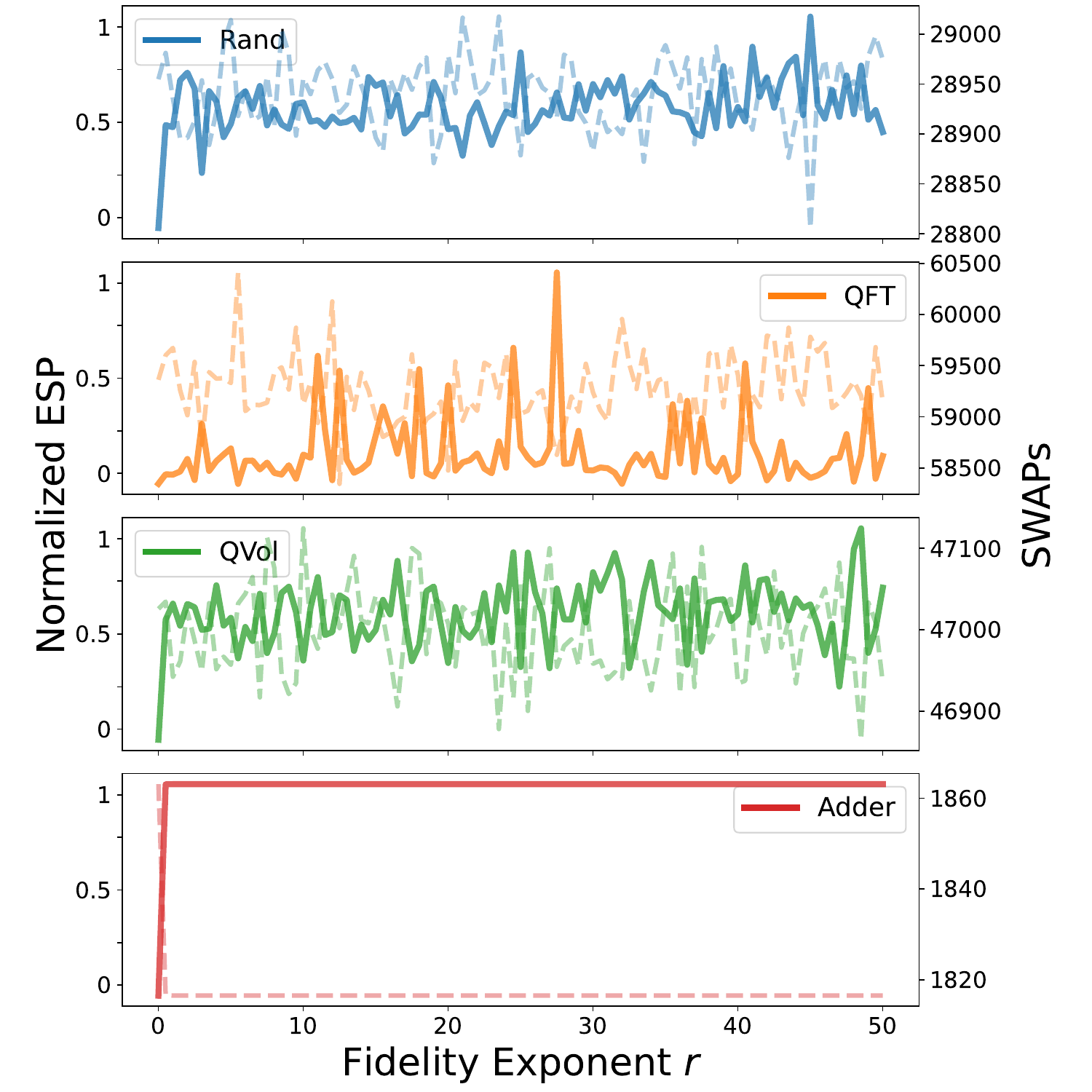}
\caption{Estimated success probability obtained after routing using Route-Forcing for an increasing value of $r$. Solid lines represent the normalized ESP (averaged over 50 trials), and the dashed lines represent the number of \texttt{SWAPs} (right axis).}
\label{fig:fidelity_r}
\end{figure}

In Figure \ref{fig:fidelity_r}, it can be seen how, for most of the selected benchmarks, a fidelity exponent of $r=0$ (i.e. not taking into account the fidelity of each link) leads to a lower ESP than when considering each link's fidelity when computing the routing. However, it can be seen that the ESP does not increase when the $r$ value is increased. The number of \texttt{SWAP} gates added is shown using the dashed plots, corresponding to the right \textit{y}-axis.

\subsection{Avoiding inter-core links}
Similar to the previous section, we evaluate how considering the link fidelity can improve the routing of the circuit, this time in the chiplet architecture setting, where qubits are distributed among cores, connected with high-fidelity links, while the cores are connected with a low fidelity link, as depicted in Figure \ref{fig:multi-core}.

We set the core size to 16, each core having a grid topology of $4 \times 4$ qubits. The inter-core edge fidelity is set to $0.98$ \cite{gold_entanglement_2021}, and, aiming at minimizing the number of inter-core edge uses, we explore the impact of $r$ on the number of uses of the inter-core links. The fidelity of all intra-core links is set to $1.0$.

We map the selected quantum circuits into a system with 16 cores (arranged in a $4 \times 4$ grid), where the inter-core topology of each core is also a $4 \times 4$ grid, resulting in a 256 qubits system (16 cores and 16 qubits per core).

\begin{figure}[t]
\centering
\includegraphics[width=\columnwidth]{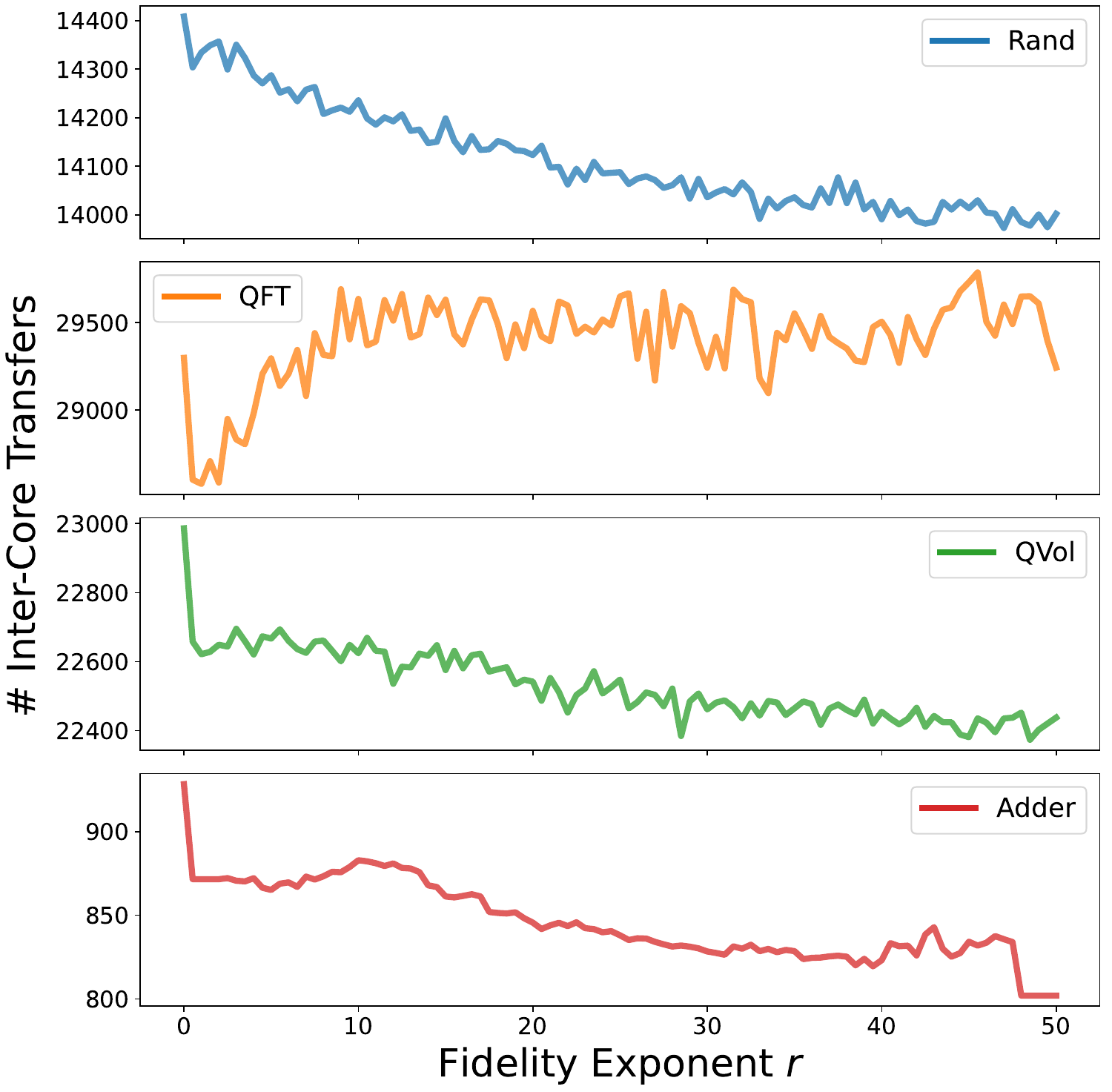}
\caption{Number of uses for the inter-core links obtained after routing using Route-Forcing for an increasing value of $r$. Results are averaged over 50 trials.}
\label{fig:inter_core_r}
\end{figure}

Figure \ref{fig:inter_core_r} depicts the number of inter-core transfers for the selected benchmarks. A higher $r$ value results in lower use of the inter-core, low-fidelity links in the chiplet architecture. These results validate the use of Route-Forcing as a mapping algorithm for modular architectures containing different types of quantum links.

\subsection{Increasing the number of qubits}
In this final experiment, we want to show how the compilation time will increase when scaling the number of qubits. Previous experiments have only reached a relatively low number of qubits (1024 in Figure \ref{fig:mapping_monolithic}). 

Figure \ref{fig:time_scaling} shows the execution time for QFT and a Random Circuit with a depth of 40 when increasing the number of qubits, from 100 qubits to 10000 ($10 \times 10$ and $100 \times 100$ grid structure respectively). The figure also shows the execution time speedup obtained when going from Python to C++.

The execution time of the Route-Forcing algorithm increases polynomically with the system size (i.e., the number of qubits). On average, for both tested circuits, the C++ implementation of the algorithm obtains a $2\times$ speedup regarding compilation time.

\begin{figure}[t]
\centering
\includegraphics[width=\columnwidth]{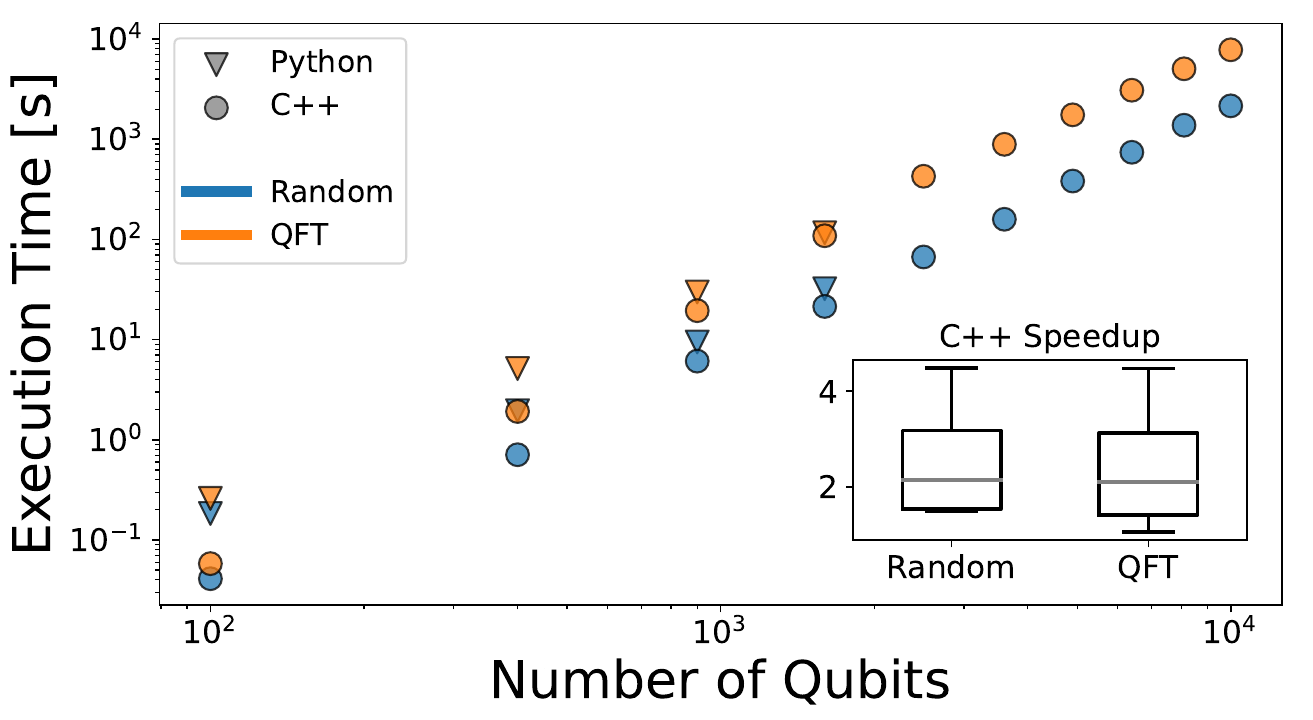}
\caption{Compilation time up to 10000 qubits for the Route-Forcing algorithm implemented in Python and C++ in a log-log scale for a Random Circuit with a depth of 40, and the QFT circuit.}
\label{fig:time_scaling}
\end{figure}

\section{Related Work}
\label{sec:related_work}


The rapidly evolving field of quantum computing has led to significant research in quantum circuit mapping and compilation. Researchers are exploring various goals, from speeding up execution times to designing systems seamlessly blending architecture and compilation. In this section, we take a closer look at different research paths that contribute to shaping the quantum mapping and compilation field. This exploration offers insights into the innovative strategies and challenges propelling advancements in quantum computing systems.

\vspace{0.3cm}
\noindent
\textbf{Improved quantum mapping.} Regarding quantum mapping research, most works have focused on reducing the number of \texttt{SWAP} operations, while less attention has been placed on reducing the depth of the resulting circuit or the compilation time. 
On the former aspect, Niu \textit{et al.} \cite{9205650} propose a hardware-aware mapping algorithm that takes into account calibration data, aiming at improving the overall fidelity of the circuit. Although their focus is not on the reduction of compilation time, their evaluation shows a $1.24\times$ compilation time speedup when compared to SABRE. However, their evaluation is limited to 20 qubits, as opposed to our assessment scaling to thousands of qubits.

On the latter aspect, Chi \textit{et al.} \cite{10.1145/3445814.3446706} propose a time-optimal \texttt{SWAP} insertion scheme. However, time optimality does not refer to a reduction in the compilation time but rather to the depth of the mapped circuit, which reduces the quantum execution time and hence improves the overall fidelity. Thanks to their approach, the authors achieve a $1.23\times$ reduction of the circuit depth compared to SABRE, but only in architectures of up to 16 qubits. Our proposed scheme improves not only the circuit depth but also the compilation time, and it is evaluated for architectures in excess of a thousand qubits. 



\vspace{0.3cm}
\noindent
\textbf{Co-design complementary approaches.} The co-design of the compilation process in general, and the mapping in particular, together with the entire architecture, has been an active area of investigation. Bandic \textit{et al.} showed the need for tight co-design among layers in a vertical cross-layer approach in \cite{bandic_full-stack_2022}, focusing on the development of hardware-aware and algorithm-driven compilation techniques. An example of this approach is given in \cite{Crawford2023}, where specific compilation considerations are provided for a 2D spin-qubit architecture to reduce the compilation overhead in terms of added \texttt{SWAP} gates. However, no mention of the scaling of the compilation time with the number of qubits is given.

A step further in considering the architecture and algorithm in the compilation process is given by Peham \textit{et al.} in \cite{optimal_subarchitectures}. The work acknowledges the exponentially increasing search space of quantum circuit mapping with the number of qubits and seeks to divide the entire architecture into optimal smaller sub-architectures so that the solution space of the mapping process is reduced. The benefit of this work is orthogonal and complementary to that of Route-Forcing since a massive quantum computer could be divided into sub-architectures, and then the mapping be sped up further thanks to Route-Forcing. 


Finally, the impact of alternative ways to route the qubits other than a \texttt{SWAP} on the quantum mapping has also been investigated. Specifically, theoretical studies have considered leveraging quantum teleportation \cite{Gottesman1999} even in single-core architectures. In this direction, Hillmich \emph{et al.} \cite{hillmich_exploiting_2020} explore how the use of quantum teleportation between distant qubits in a single core can help significantly reduce the overhead of quantum mapping in terms of a number of added gates. In Route-Forcing, the formulation of the attraction forces could be modified so that such long-range communication primitives can also be taken into consideration.

\vspace{0.3cm}
\noindent
\textbf{Mapping for multi-core and distributed architectures.} The mapping problem extends its scope to multi-core and distributed quantum computing architectures. In the context of multi-core quantum processors, approaches based on partitioning of the qubit interaction graph on a timeslice basis have been investigated \cite{escofet_hungarian_2023, escofet_revisiting_2024, bandic_mapping_2023, baker_time-sliced_2020}. These works aim to minimize the number of inter-core qubit state transfers in the pathway of maximizing the overall fidelity. However, both algorithms assume full qubit connectivity inside and across the quantum processors to reduce the problem to a graph partitioning problem. Hence, contrary to Route-Forcing, these algorithms are not directly applicable to monolithic (single-core) architectures or architectures with real topologies.

Regarding quantum circuit mapping for distributed networks, research has delved into compiler strategies, as in \cite{ferrari_compiler_2021}, where the authors discuss the main challenges arising with compiler design for distributed quantum computing and derive an upper bound of the overhead induced by quantum compilation. Later, in \cite{ferrari_modular_2023}, Ferrari \textit{et al.} propose a quantum compiler strategy for distributed quantum computing, considering both network and device constraints. The huge cost of qubit state transfers in this scenario renders the proposed solutions inadequate for single-core and multi-core quantum computers and hence not comparable to Route-Forcing.



\section{Conclusions and Future Work}
\label{sec:conclusions}

Quantum computers are needed to scale to close the gap that currently exists between quantum algorithms and quantum hardware. This work proposes a novel mapping algorithm focusing on the future of quantum computing architectures, obtaining a fast solution and adhering to the constraints that will come with multi-core architectures.

The Route-Forcing algorithm outperforms the state-of-the-art alternative in terms of the depth of the resulting mapped circuit and execution time, making a remarkable contribution to quantum circuit mapping for both single-core and multi-core architectures.

Moreover, the exhaustive space exploration conducted shows the potential to tune the proposed mapping algorithm to optimize a particular metric. This allows the end user to prioritize depth, time, or the number of added \texttt{SWAP} gates by modifying the $p$ and $k$ parameters.

Some future lines of research include exploring other $\texttt{SWAP}$ policies, tunning the attraction force, the interaction weighting factor, and, in the multi-core scenario, the exponent of the fidelity. Moreover, other types of quantum communications different than $\texttt{SWAP}$ gates can be considered.

When the architecture has more available qubits than the circuit uses, the unused qubits of the architecture can be used as communication resources. By entangling them and driving them apart, they would allow for EPR-based communications such as quantum teleportation or remote gate execution \cite{Gottesman1999}, reducing the depth even further.

Quantum circuit mapping and quantum compilation, in general, need to adapt to the growing size of quantum computing architectures. This work helps close the gap between quantum algorithms and quantum hardware in the years to come, paving the way to unleashing quantum computing's full potential.

\section*{Acknowledgment}
Authors gratefully acknowledge funding from the European Commission through HORIZON-EIC-2022-PATHFINDEROPEN-01-101099697 (QUADRATURE) and grant HORIZON-ERC-2021-101042080 (WINC). This work has been partially supported by the Spanish Ministry of Science, Innovation and Universities through the Beatriz Galindo program 2020 (BG20-00023), by the  European ERDF under grant PID2021-123627OB-C51, and by QCOMM-CAT-Planes Complementarios: Comunicacion Cuántica
supported by MICIN with funding from the European Union, NextGenerationEU (PRTR-C17.I1) and Generalitat de Catalunya.  The first author gratefully acknowledges the Universitat Politècnica de Catalunya and Banco Santander for the financial support of his predoctoral FPI-UPC grant.

\bibliographystyle{IEEEtran}
\bibliography{main}

\end{document}